# Omni-iEEG Planner: an open-source Python platform integrating multimodal neuroimaging and intracranial EEG data for epilepsy surgery research

Yuta Tanoue,[1]* Masaki Izumi,[1] Yipeng Zhang,[2] Chenda Duan,[2,3] Tonmoy Monsoor,[3] Christopher A. Chan,[5] Joe X Qiao,[4] Noriko Salamon,[4] Aria Fallah,[5] Vwani Roychowdhury,[3] and Hiroki Nariai[1]*

1. Division of Pediatric Neurology, Department of Pediatrics, David Geffen School of Medicine at UCLA and UCLA Mattel Children's Hospital, Los Angeles, CA, USA
2. Google Inc. Mountain View, CA USA.
3. Department of Electrical and Computer Engineering, University of California, Los Angeles, CA, USA.
4. Department of Radiology, UCLA Medical Center, David Geffen School of Medicine, Los Angeles, CA, USA.
5. Department of Neurosurgery, David Geffen School of Medicine at the University of California Los Angeles, Los Angeles, CA, USA.

*Corresponding author: Yuta Tanoue, MD, PhD (yu.tanoue0712@gmail.com); Hiroki Nariai, MD, PhD (hnariai@mednet.ucla.edu)

## Abstract

Multimodal intracranial EEG (iEEG) research for epilepsy surgery combines anatomical and functional imaging, implanted electrode locations, interictal biomarkers, and functional mapping—outputs produced by different applications and reconciled by hand before they can be viewed on the same anatomy. Omni-iEEG Planner is an open-source Python platform that brings these inputs into one patient coordinate frame for research data analysis. The workflow starts from the preoperative T1-weighted MRI, from which FreeSurfer provides the cortical surface and anatomical labels; CT, PET and diffusion MRI are then co-registered to that reference, electrode contacts are detected automatically on the postimplantation CT and reviewed by the analyst, and tractography is reconstructed within the application. Channel-indexed results computed outside the platform, such as high-frequency oscillation counts from pyHFO or time-resolved gamma responses, are projected onto the contacts and the cortical surface with their source units preserved, and interictal and functional evidence remain separate layers. We describe the design and demonstrate representative workflows on de-identified patient data. The platform is not proposed as a method for every laboratory; it is a worked example of how one research group combined multimodal integration with an AI-assisted iEEG analysis tool, pyHFO, in a single open environment that others can adapt to their own pipelines. The design will accommodate further evidence types—model-derived contact scores, virtual implantation and hypothetical resection studies, group-level analyses—as they are implemented. The software is released under the MIT license for research and educational use, not clinical decision-making.

## 1. Introduction

Presurgical evaluation of drug-resistant focal epilepsy with intracranial EEG (iEEG) produces several kinds of spatial evidence about the same brain: structural MRI and the cortical surfaces and labels derived from it; PET and other metabolic or functional images; diffusion MRI and tractography; the coordinates of implanted subdural and depth contacts; interictal biomarkers computed from the recordings, such as spikes and high-frequency oscillations (HFOs) [1,2]; functional maps from task-related high-gamma activity [3]; and clinically annotated contact-level findings—the seizure onset zone and the results of cortical stimulation mapping [4]—which are recorded by hand for each contact. Each is produced by a different tool with its own file formats, coordinate conventions and channel naming. In research practice the pieces are brought together manually: electrode names are reconciled between the localization software and the EEG analysis, images are re-registered in a viewer, annotations are transcribed from clinical reports, and the result is a figure that cannot be queried again. The work is repeated for every patient and again for every new biomarker.

Clinical planning systems used at our center—a commercial cranial planning platform and the ROSA robotic workflow for stereo-EEG implantation [5]—are designed for image-guided surgery and are proprietary. In our workflow, imaging-derived findings such as PET hypometabolism or fMRI activation are prepared in separate applications and transferred into the planning system as targets, and quantitative iEEG results are not part of the planning display. These clinical systems are not the place to prototype a new biomarker, and their code cannot be inspected or modified. That is the gap a research platform fills.

A substantial body of open research software addresses parts of this problem, and Omni-iEEG Planner builds on it. Brainstorm [6] provides MEG, EEG and iEEG analysis with MRI-, CT- and PET-based anatomy and contact localization within MATLAB; 3D Slicer [7] is a general image-computing platform with iEEG extensions [8]; MMVT [9] renders MRI, CT, PET, fMRI, MEG/EEG and electrode data in Blender; and IntrAnat Electrodes [10] combines SEEG localization, PET, fMRI and diffusion display, functional labeling of contacts and a group database within BrainVISA. Dedicated localization tools—iELVis [11], the FieldTrip iEEG pipeline [12], the MNE-Python localization interface [13], LeGUI [14], iElectrodes [15], EpiTools and GARDEL [16], img_pipe [17], SEEG Assistant [8], ALICE [18] and iEEGview [19]—solve contact detection and anatomical labeling, largely through rigid CT–MRI registration and FreeSurfer surfaces [20–23], following the approaches introduced by Hermes [24], Dykstra [25] and Arnulfo [26]. Planning platforms such as EpiNav address trajectory design [27]. Omni-iEEG Planner was designed specifically for our group's multimodal workflow rather than as a replacement for any of these tools: it connects FreeSurfer anatomy, CT-based contact localization, pyHFO biomarker analysis [28,29] and Detroit-procedure high-gamma mapping [3,30]—the components we already use—inside one Python environment.

Four requirements shaped the design. First, one patient coordinate frame: every image, surface, contact and result is expressed in the patient's own T1 space, so that nothing has to be re-registered to be compared. Second, result-first import: any table indexed by contact or channel pair—an HFO count, a gamma time series, a clinical annotation, a future model score—can be loaded and rendered without code changes, with its units and analysis interval carried from the source. Third, traceable layers: interictal biomarkers, functional responses and clinical annotations can be viewed together on the same map, but each remains an independently controlled layer with its own scale, units and provenance rather than being blended into a single composite score, so that a researcher can always see which evidence produced a given color at a given contact. Fourth, a Python-native implementation. Brainstorm accomplishes much of the anatomical integration described here within MATLAB, 3D Slicer exposes Python interfaces, and the multimodal display concept resembles that of MMVT; the Planner differs in being built from the start to work seamlessly with deep-learning tools. Our biomarker software (pyHFO [28,29]) is Python-based, so the platform exchanges arrays and tables with it directly, can be scripted for batch processing, and is installed with the same package tools as the analyses themselves.

This report describes the design and the implemented functions of Omni-iEEG Planner (Prototype 1.0 RC20) and demonstrates them on de-identified patient data. It does not report quantitative accuracy of registration or localization, comparisons with other software, or clinical outcomes; those are subsequent studies. Nor do we propose the Planner as a method for every laboratory. In the era of AI-assisted coding, each research team is likely to assemble and customize the tools that fit its own data and questions; what we present is one team's worked example of integrating a multimodal platform with an AI-assisted iEEG analysis pipeline, offered so that others can see how it was done and adapt it. Omni-iEEG Planner is intended for research and educational use and is not intended for clinical decision-making or the execution of surgical procedures.

## 2. Software design and functionality

### 2.1 Architecture, data model and development

Omni-iEEG Planner (V1.0 RC20; software version 1.0.0rc20) is a desktop application implemented in Python (Python 3.12.14, 64-bit) with a graphical interface built using PySide6 and three-dimensional rendering through VTK [31]. Images are read and written with NiBabel and SimpleITK, pydicom is used to inspect DICOM metadata, and numerical operations use NumPy [32] and SciPy [33]. The documented development and test environment is Windows 11 Enterprise, 64-bit. Each patient is represented by a project folder containing a JSON manifest and associated image and numerical assets for the reference MRI, registered images, loaded FreeSurfer-derived data, reconstructed contacts and imported results. Saving and reopening a project restores image layers, regions of interest, contacts and their identifiers and review status, imported analyses, saved tractography and recorded display settings, images, surfaces and contacts are displayed in a common patient-specific scanner RAS coordinate frame, in millimeters, while native image geometry and registration transforms are retained. The coordinate convention is shown in the status bar; no template normalization is applied in the reported workflow. The tabs used for the workflows described—Viewer, Segmentation, Analysis and DTI/Tracts–organize image review, anatomical processing, analysis display and tractography (Figure 1). Development used AI-assisted coding and debugging with ChatGPT and Codex (GPT-6 Astra). Development verification included synthetic-data tests, regression checks and manual checks of graphical-interface behavior. Dependencies and versions are listed in Table 2.

### 2.2 Image import and co-registration

Images are imported as DICOM series or NIfTI volumes. The preoperative T1-weighted MRI is the reference; in the demonstration data, the reference volume had 1.0-mm isotropic voxels in a 256 × 256 × 192 matrix. Outputs generated outside the application using FreeSurfer version 8.2.0 [20–22]—the pial surfaces and the aparc+aseg volume carrying Desikan–Killiany cortical labels [23] and subcortical

segmentations—are loaded from the subject directory and supply the cortical surface, anatomical labels and hover-over region. Additional image volumes, such as other MRI sequences, postimplantation CT, contrast-enhanced CT (including CT angiography/venography), and PET can be registered to the reference using six-degree-of-freedom rigid registration in SimpleITK [34], with geometry-based initialization, multiple candidate starting orientations, a Mattes mutual-information metric and Powell optimization, following a mutual-information approach [35–37]. Registered images appear as layers in three orthogonal slice views and in the 3D view; each layer's visibility, opacity, window and color map are set independently, and a compare mode toggles the reference and the registered image for visual inspection of alignment. PET is displayed as a color overlay on the MRI with an adjustable intensity range (Figure 2B). The application displays the PET volume as acquired; hypometabolism maps or statistical parametric images are upstream analyses and are imported, when available, as additional volumes.

## 2.3 Anatomical regions, ROIs and tractography

Regions are selected from the FreeSurfer label list (156 labels in the demonstration subject) or drawn manually on the slices with a brush of adjustable radius. A selected region is outlined on the slices, rendered in the 3D view with the electrodes, and can serve as a seed or filter for tractography (Figure 2C). Diffusion MRI is imported as a DICOM series converted using dcm2niix [38], or as a 4D NIfTI volume accompanied by b-value and b-vector files. The demonstration acquisition comprised 25 volumes—one b = 0 volume and diffusion-weighted volumes at b = 600 and 1200 s/mm$^2$—with a 128 × 128 × 40 matrix and 1.875 × 1.875 × 4.0 mm voxels. Motion correction is applied on import by rigidly registering the diffusion-weighted volumes to the b = 0 reference and reorienting the gradient vectors accordingly; no echo-planar distortion correction is performed within the application. The diffusion tensor and fractional anisotropy (FA) are reconstructed using a weighted-least-squares single-tensor model implemented in DIPY version 1.12.1 [39], and deterministic tractography is performed from selected ROIs. Streamlines passing through one or two selected ROIs are retained and displayed together with the anatomy and the contacts (Figure 2D). Direction-encoded FA can be overlaid on the slices, and the b = 0 image can be compared with the T1-weighted MRI to inspect alignment.

## 2.4 Localization and review of implanted contacts

The postimplantation CT is rigidly registered to the reference MRI using the method described in Section 2.2. Candidate contacts are detected by high-attenuation voxels corresponding to electrode metal in the native CT. The analyst reviews the detected leads and can create additional leads semiautomatically by specifying the tip contact, last contact and total contact count. The review dialog (Figure 3) shows two reformatted CT planes along the selected lead, a cross-section through the selected contact, and a whole-brain 3D view. A table lists every contact with its status, peak CT attenuation and spacing from the previous contact. The analyst can rename a lead, reverse its tip and end, snap a contact to the local metal center, add or delete contacts, and confirm the lead. Confirmed contacts carry the identifiers used by later analyses. Monopolar and bipolar labels can be edited to match external tables; a bipolar pair is displayed at the midpoint of its two contacts and keeps its pair identity in every table and map.

## 2.5 Import and mapping of interictal biomarkers

Results computed outside the application are imported as Excel (.xlsx) tables with one row per contact or channel pair and one or more value columns. Identifiers are matched to the reconstructed contacts; the mapping dialog reports matched and unmatched channels and lets the analyst reconcile naming differences. Imported values are preserved, and units can be specified for each metric. In the illustrated workflow the values are spkHFO event counts from pyHFO [28,29] over the analyzed interval; the application does not convert counts to rates and does not combine results from different sources. Values are shown in a channel-indexed table with color bars and on the cortical surface as a heatmap generated by distance-weighted interpolation from contact positions or bipolar midpoints to nearby surface vertices within a user-set distance (15 mm in the examples). An optional display threshold controls which values are shown, and the color range is set manually or from the data. The same route supports other channel-indexed interictal biomarkers supplied in a compatible table format, including outputs of the deep-learning HFO classifiers in pyHFO [40–42] and biomarkers shown to predict surgical outcome [43].

## 2.6 Functional mapping and time-resolved display

Functional results are imported and displayed separately from interictal biomarkers. Task-related high-gamma responses from the Detroit procedure [3,30,44], computed outside the application using BESA [45], are imported as a channel-by-time table. In the demonstration the table held 167 channels × 261 time points spanning approximately −600 to 2000 ms in 10-ms steps relative to stimulus onset. A time slider and playback controls select the time point; the table and the cortical map update together (Figure 4C–D), and the sequence can be exported as a video. High-gamma mapping has been compared with cortical stimulation mapping, the reference standard for functional localization [4,46–50]; direct entry of stimulation findings into the same workspace is planned (Section 4.3).

## 2.7 Display, export and project save

Slice and 3D views can be arranged as one, four or six panels, and the 3D view is rotated, zoomed and reset to standard orientations. Still images of the current view and time-resolved sequences are exported as PNG and MP4 files, respectively. Contact tables containing anatomical labels and coordinates are exported as CSV files, and contact-coordinate records are exported as JSON files for downstream analysis. Coordinates are expressed in patient-specific scanner RAS (right–anterior–superior) coordinates in millimeters, with positions in the reference MRI and, when available, the native CT. Projects are saved and reopened with their data and display settings as described in Section 2.1. Removing an image layer does not delete the underlying DICOM or NIfTI files.

## 2.8 Demonstration data and ethics

Demonstrations use images and analysis results from 3 patients who underwent iEEG monitoring for drug-resistant epilepsy at UCLA Mattel Children's Hospital, acquired during routine clinical care and analyzed retrospectively under UCLA IRB protocol #18-1599, which waived the requirement for informed consent. No examination or procedure was performed for this report. All displays are de-identified. The imaging examples in Figure 2A–B come from one patient, the region and tractography examples in Figure 2C–D and the contact review in Figure 3 from a second, and the biomarker and functional examples in Figure 4 from a third.

## 3. Representative workflows

### 3.1 Constructing a multimodal patient model

The analyst opens a patient project, imports the T1-weighted MRI and the FreeSurfer subject directory, and sees the reference volume in three orthogonal slices with the reconstructed cortex in the 3D panel (Figure 2A). Hovering over a slice reports the FreeSurfer label under the cursor. Adding the PET series registers it to the reference and shows it as a color overlay whose opacity and intensity range are adjusted on the left panel (Figure 2B). In the Segmentation tab the analyst searches the label list and selects the left thalamus and left hippocampus, which are rendered as colored regions inside the translucent cortex together with the reconstructed depth electrodes (Figure 2C). In the DTI/Tracts tab the diffusion series is imported, the two regions are chosen as ROIs, and the streamlines passing through both are displayed with the cortex and the contacts (Figure 2D). Every element in Figure 2 is in the same coordinate frame, so a contact can be judged against the PET signal, the parcel it lies in, and the tracts it neighbors without leaving the view.

### 3.2 Reviewing implanted contacts

After the postimplantation CT is imported and registered, the Postoperative review tab lists the detected leads. Selecting a lead opens the review dialog (Figure 3): the two reformatted planes along the lead show the contacts as bright discs on the electrode shaft, the cross-section confirms the selected contact, and the table lists the eight contacts with their peak attenuation and inter-contact spacing. Spacing values close to the electrode's nominal pitch indicate a consistent reconstruction along the shaft; values that depart from it, or contacts flagged as possible duplicates, prompt the analyst to snap, move or delete a contact before confirming. Confirmed leads and their identifiers are then available to every result imported later.

### 3.3 Mapping interictal and functional results on the same cortex

In the Analysis tab the analyst imports the pyHFO result table for the same patient. The mapping dialog matches all 167 bipolar channels to reconstructed contacts, and the static-results view shows each channel's spkHFO count next to a color bar and, in the 3D panel, as a heatmap on the cortex (Figure 4A–B). In the demonstration, counts ranged from 0 to 418 events per channel; the highest values were in the left anterior hippocampal and left amygdala leads and in the right hippocampal and posterior temporal leads. The analyst then imports the gamma table for the same contacts. The time-series view adds a slider; moving it updates the channel table and the cortical map to the selected latency (Figure 4C–D), and the sequence can be played or written to a video. The two results are independent layers: they share the contacts and the cortex but keep their own units, color scales and thresholds, so an interictal count and a functional response are compared by position rather than blended into one image.

### 3.4 Relation to existing tools

Table 1 summarizes the functions of Omni-iEEG Planner alongside those of the platforms most often used for the same tasks, as described in their publications and documentation. The overlap is substantial and intended: contact localization on CT, FreeSurfer anatomy and multimodal overlay are mature methods and are implemented here in the same way. The Planner differs in the combination it packages: in-application tractography, a generic import of channel-indexed result tables with an optional time axis, separate interictal and functional layers, and a Python implementation that shares its language with pyHFO and PyTorch. IntrAnat Electrodes is the closest relative in scope; it is built on BrainVISA, labels contacts with functional annotations and supports group studies, whereas the Planner emphasizes direct display of externally computed result tables and time-resolved maps. Group analysis is on the Planner roadmap (Section 4.3).

## 4. Discussion

### 4.1 What the platform contributes

Omni-iEEG Planner brings the imaging, contact reconstruction and iEEG results of one patient into a single coordinate frame and lets an externally computed result be placed on that anatomy without conversion. The demonstrated workflows move from images to reviewed contacts to biomarker and functional maps in one application, and every displayed value can be traced to its source table, unit and channel identifier. The platform was built around the methods our group already uses; its value to other centers lies in the same property—it is a documented, modifiable implementation that can be adapted to a different biomarker, functional paradigm or analysis pipeline rather than a closed product.

The relation to existing software is complementary. Brainstorm remains the reference environment for electrophysiological analysis with anatomy [6]; MMVT and 3D Slicer offer broader visualization and image-processing capabilities [7,9]; IntrAnat Electrodes provides functional contact labeling and a group database [10]. The Planner's contribution is narrower: a result-first workspace in which any channel-indexed table—including outputs of deep-learning models—lands on the patient's own cortex with its meaning intact. That design also determines what the platform does not do. It does not compute biomarkers, activation maps or hypometabolism; it displays what upstream analyses produce, and their limitations travel with them.

### 4.2 Python and interoperability

The choice of Python follows from the pipeline it serves. pyHFO [28,29], the HFO classifiers derived from it [40–42] and our developmental and network analyses [51,52] are Python code; PyTorch [53] models produce contact-level scores as arrays. A Python application reads those outputs directly, can be driven from a script to process a cohort, and is installed with the same package tools as the analyses themselves. The practical consequence for a research group is that a new biomarker reaches the anatomical display as soon as it exists as a table. This does not make the Planner faster or more capable than MATLAB-based tools, and we make no such claim; Brainstorm runs as a compiled standalone application, and MMVT and 3D Slicer also provide Python access. The benefit is the absence of a language boundary in one specific workflow.

### 4.3 Development roadmap

Several extensions follow from the present design and are planned rather than implemented. Virtual implantation will represent hypothetical SEEG trajectories from user-selected entry and target points with a specified electrode geometry, displayed with the anatomy and existing contacts and kept distinct from reconstructed implants. A hypothetical resection or ablation workflow will let a researcher save a selection of contacts or a volume, view the biomarker and functional layers that fall inside it, and record the rationale; such saved selections are the substrate for studies of how disclosing a biomarker map changes a plan. Exchange of plans with the ROSA workflow is an intended interoperability path that will require a defined export format, a specified receiving software version, and verification that entry and target coordinates, electrode identity and the reference image are preserved; no such transfer has been tested. Manual entry of stimulation-mapping findings—the stimulated pair, the task, the observed response and the stimulation parameters, with untested sites distinguished from sites tested without response—will add the third class of evidence to the workspace. Execution of pyHFO within the application, so that detection, classification and display occur without leaving the workspace, is planned on the same table interface. AI-derived contact-level scores will be imported with their model version and source recording so that a map can be traced to the model that produced it. Group-level aggregation of contacts across patients, with each contact carrying its patient identity, native coordinates, a defined common-space coordinate and an atlas label, will support cohort questions such as the anatomical distribution of thalamic SEEG contacts; the developmental atlases of physiological HFOs and phase-amplitude coupling [51,52], the tractography atlases of language networks [54–56], normative iEEG maps [57,58] and our thalamic and neuromodulation studies [59–62] are the kinds of analysis this would serve. Structured exports will follow iEEG-BIDS conventions [63] for electrode and recording metadata, and the aligned imaging–iEEG representation is a natural substrate for multimodal models building on the Omni-iEEG dataset [64].

### 4.4 Limitations

This report describes development and demonstration, not validation. Registration and contact-localization accuracy, reproducibility after export and reload, behavior across acquisition protocols and vendors, and usability have not been measured. The examples come from a small number of patients at one center, and the software is a prototype whose interface and file formats may change. The cortical heatmap is a display transformation: a projected color can extend beyond the tissue that was sampled, and bipolar values refer to a pair of contacts rather than a point. Imported biomarkers and functional responses retain the limitations of the analyses that produced them, and the application does not validate a result by displaying it. Code produced with AI assistance was reviewed by the developers, but coordinate handling, numerical operations and exports require the same independent verification as any research software. Our group has 97 patients previously processed in Brainstorm; a consistency study comparing contact coordinates and surface projections between the two platforms, with matched coordinate conventions and documented registration differences, is the next technical step, and those cases are not a validation cohort in this report.

## 5. Software availability and conclusion

The Omni-iEEG Planner source code (Prototype 1.0 RC20) is available at [https://github.com/yutanoue0712-bit/Omni_iEEG_Planner](https://github.com/yutanoue0712-bit/Omni_iEEG_Planner) under the MIT license, and the release described here is archived at Zenodo ([https://zenodo.org/records/23028674](https://zenodo.org/records/23028674); doi:10.5281/zenodo.23028674). The release includes installation instructions and documentation of input and export formats; a demonstration project built from permitted data and a short tutorial are planned. Omni-iEEG Planner is intended for research and educational use and is not intended for clinical decision-making or the execution of surgical procedures.

Omni-iEEG Planner integrates multimodal imaging, electrode review and externally computed iEEG results in a Python research platform. Its demonstrated workflows connect a patient's anatomy with interictal biomarkers and functional mapping on the same cortex; its roadmap adds virtual planning, stimulation annotations, group analysis and AI-derived scores on the same foundation.

## Data availability

Patient imaging and electrophysiological data cannot be shared publicly. The displays in this report were generated from de-identified data and contain no identifying information.

## Funding

Research reported in this publication was supported by the National Institute of Neurological Disorders and Stroke of the National Institutes of Health under award number K23NS128318. The content is solely the responsibility of the authors and does not necessarily represent the official views of the National Institutes of Health.

## Competing interests

The authors declare no competing interests.

## Author contributions

YT: software, methodology, investigation, visualization, writing—original draft. MI: investigation, validation, writing—review and editing. YZ, CD, TM, CC, JQ: investigation, validation and review. NS, AF, VR: validation, supervision and review. HN: conceptualization, supervision, funding acquisition, resources, writing—review and editing.

## Figure legends

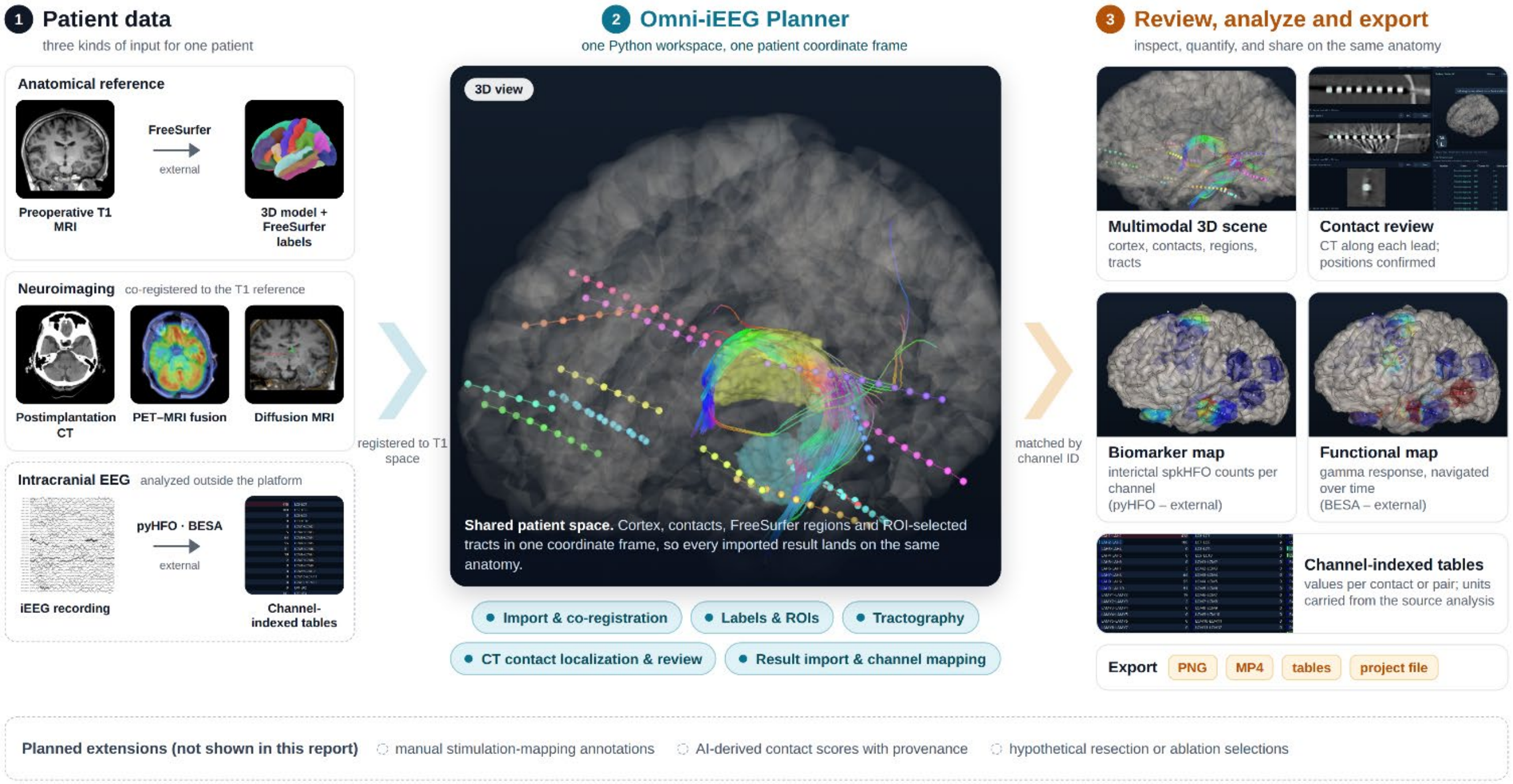

**Figure 1. Overview of the Omni-iEEG Planner workflow.** Left, patient data enter as three kinds of input. The anatomical reference is the preoperative T1-weighted MRI, processed outside the platform with FreeSurfer to produce the three-dimensional cortical model and parcellation labels that define the patient's coordinate space. Neuroimaging volumes—postimplantation CT, PET and diffusion MRI—are co-registered to that reference within the platform, and electrode contacts are detected from the CT. Intracranial EEG is analyzed outside the platform (for example with pyHFO or BESA) and enters as channel-indexed result tables. Center, all inputs share one patient coordinate frame; the chips name the implemented modules. Right, results are reviewed, quantified and exported on the same anatomy. The dashed band lists planned extensions not demonstrated in this report. The head-CT tile is a public de-identified example image and the iEEG trace tile is a schematic; all other tiles are taken from the application's own displays.

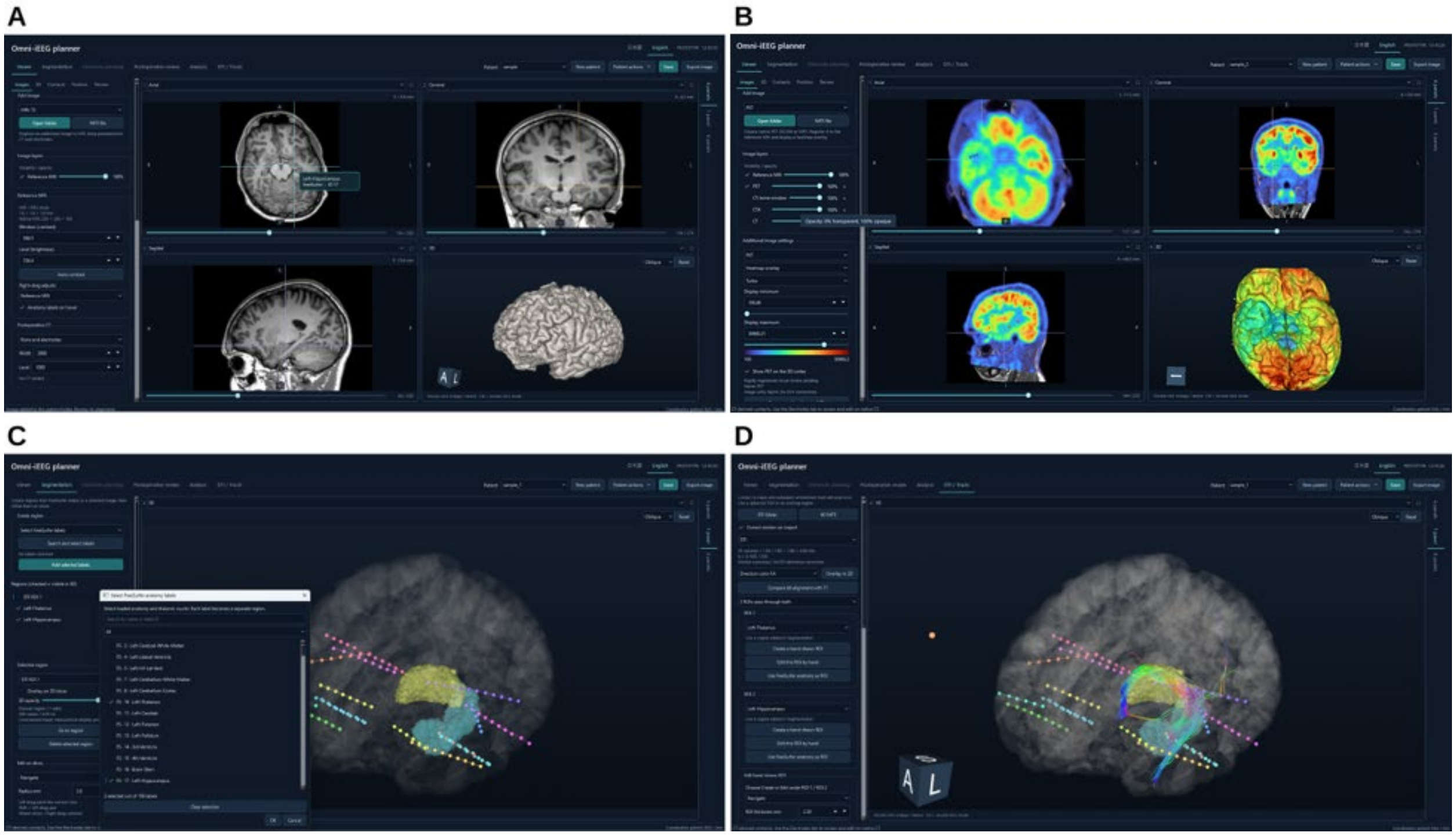

**Figure 2. Multimodal anatomy and tractography in the patient workspace.** (A) Reference T1-weighted MRI in axial, coronal and sagittal views with the reconstructed cortical surface in the 3D panel; hovering over a slice reports the FreeSurfer label under the cursor. (B) PET registered to the reference MRI and displayed as a color overlay on the three slice views and on the cortical surface in the 3D panel, with an adjustable intensity range; hypometabolism of the right temporal lobe is visible on both the slices and the cortical surface. The layer list at left shows the other registered volumes of this patient, each toggled and windowed independently. (C) Left thalamus and left hippocampus selected from the FreeSurfer label list (selection dialog shown) and rendered as colored regions inside the translucent cortex together with the reconstructed depth electrodes. (D) Streamlines passing through both regions, generated in the DTI/Tracts panel, displayed with the cortex, regions and contacts. Panels A–B and C–D come from different patients. Colors identify image intensities, anatomical objects or tracts according to each layer's settings.

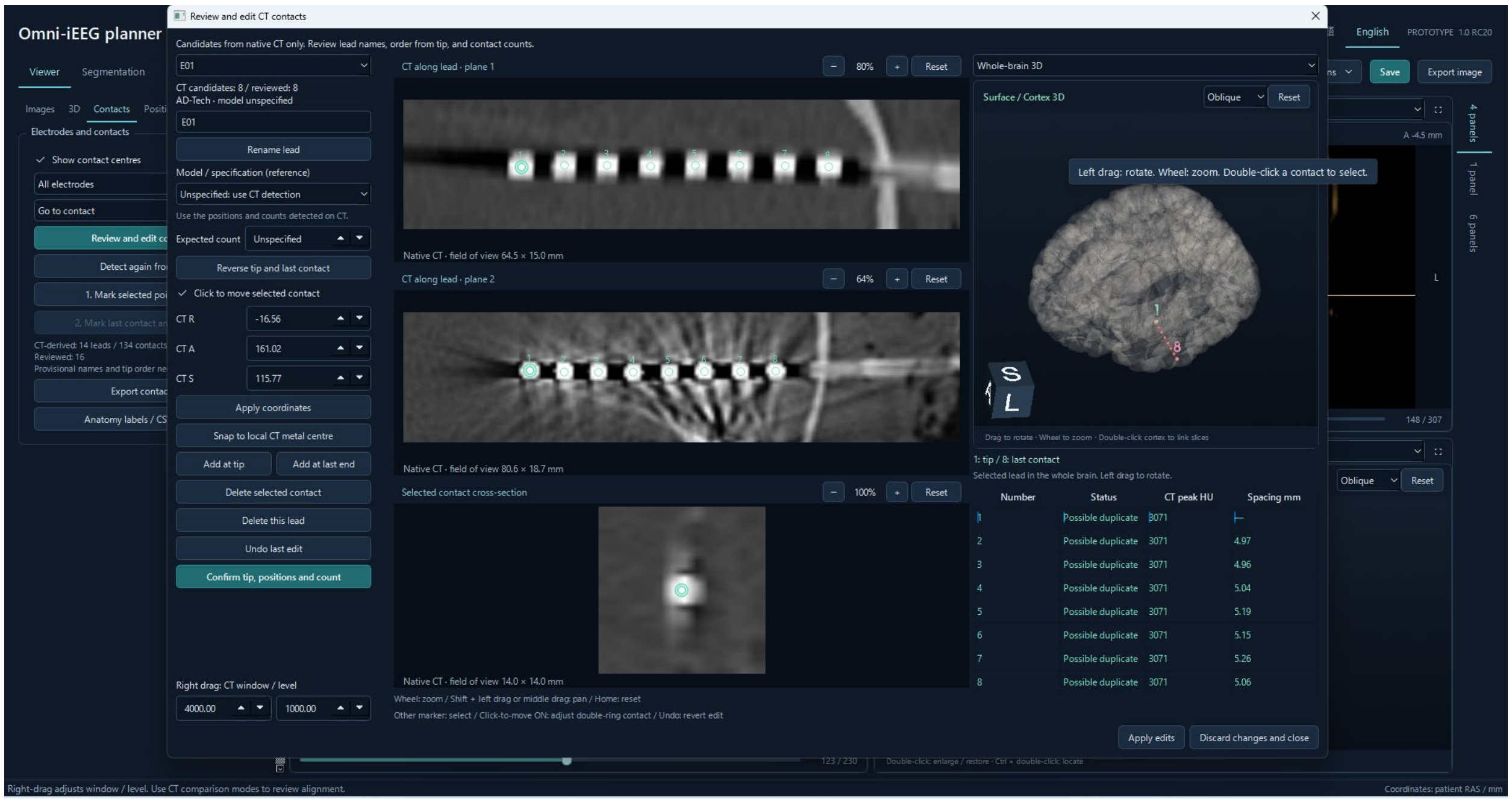

**Figure 3. Interactive CT-based review of implanted contacts.** The review dialog for one lead (E01, eight candidate contacts) shows two reformatted CT planes along the lead, in which the contacts appear as bright discs on the electrode shaft; a cross-section through the selected contact; and a whole-brain 3D view with the selected lead highlighted. The table lists each contact with its review status, peak CT attenuation and spacing from the previous contact; here the spacing of 4.96–5.26 mm matches the electrode's nominal 5-mm pitch. Controls at left rename the lead, reverse its tip and end, set the expected contact count, snap a contact to the local metal center, add or delete contacts, and confirm the tip, positions and count. The status flags ("possible duplicate") are candidate flags shown during review, not a final localization; the panel behind the dialog reports 14 CT-derived leads with 134 contacts for this patient.

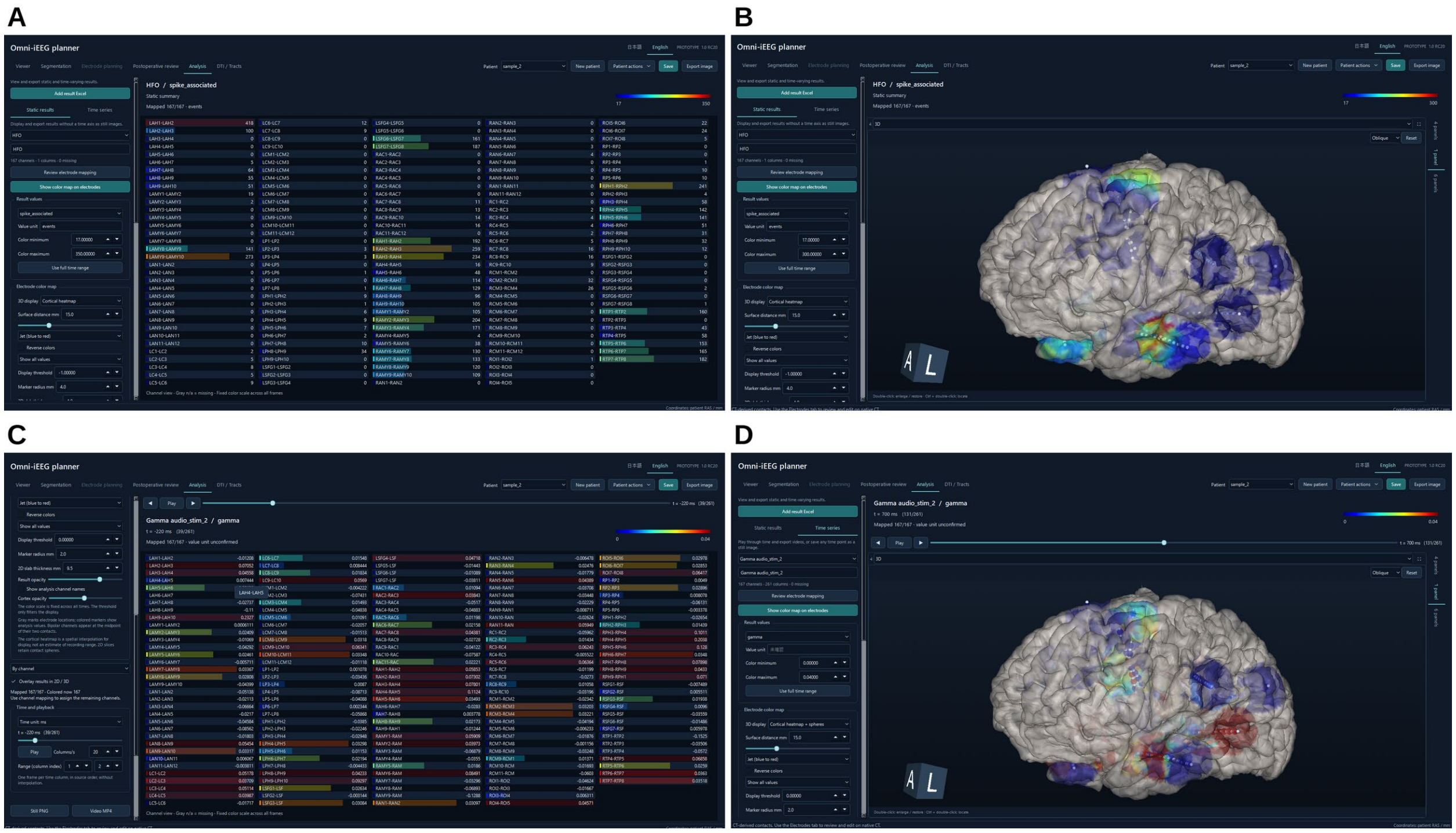

**Figure 4. Interictal biomarker and functional maps on the same cortex.** (A) Static-results view of spike-associated HFO (spkHFO) counts imported from pyHFO: all 167 bipolar channels were matched to reconstructed contacts, and each channel's count is shown with a color bar on the shared scale (17–350 events in this display). (B) The same counts projected onto the cortical surface as a heatmap; the lateral temporal region carries the highest values in this patient. (C) Time-series view of task-related high-gamma values imported as a channel-by-time table (167 channels × 261 time points); the slider is set to −220 ms relative to stimulus onset, and the table updates to that latency. (D) The cortical map of the same result at 700 ms, with contacts shown as spheres. The two results share the contacts and the cortex but are separate layers with their own units, color scales and thresholds; a color in one map cannot be read against the scale of the other.

## Tables

**Table 1**. Functions of Omni-iEEG Planner and of related open research platforms, as described in their publications and documentation (September 2026).

| Platform | Environment | CT-based contact localization with review | MRI + CT + PET overlay | Tractography in application | Import of channel-indexed result tables | Time-resolved channel maps | Group analysis | License |
|---|---|---|---|---|---|---|---|---|
| **Omni-iEEG Planner** | Python | ● | ● | ● | ● | ● | — (planned) | MIT |
| Brainstorm [6] | MATLAB; compiled standalone | ● | ● | ○ | ○ | ● | ● | GPL-3.0 |
| 3D Slicer [7,8] | C++/Python | ○ (SEEG Assistant) | ● | ○ (SlicerDMRI) | ○ | ○ | — | BSD-style |
| MMVT [9] | Python + Blender | ○ | ● | — | ● | ● | — | open source |
| IntrAnat Electrodes [10] | Python/BrainVISA | ● | ● | ○ (displays external tracts) | ○ (functional labels) | — | ● | GPL |
| iELVis [11] | MATLAB | ● | ○ | — | ● | — | ○ | open source |
| FieldTrip [12] | MATLAB | ● | ○ | — | ● | ● | ● | GPL |
| MNE-Python [13] | Python | ● | ○ | — | ● | ● | ● | BSD-3 |
| LeGUI [14] | MATLAB | ● | ○ | — | — | — | ○ | open source |

● documented in the core application; ○ available through an extension, an external pipeline or manual steps; — not described in the cited sources. Absence of a mark means we did not find the function documented, not that it cannot be achieved.

**Table 2.** Software environment and parameters of Omni-iEEG Planner Prototype 1.0 RC20.

| **Component** | **Value** |
|---|---|
| Python version | CPython 3.12.14, 64-bit. |
| Graphical interface | PySide6_Essentials 6.11.2 and shiboken6 6.11.2. |
| 3D rendering | VTK 9.7.0. |
| Numerical processing | NumPy 2.5.3 and SciPy 1.18.1 |
| Image I/O | NiBabel 5.4.2 for NIfTI and FreeSurfer images and surfaces; SimpleITK 2.5.6 for DICOM series reading, registration and resampling; pydicom 3.0.2 for DICOM header inspection |
| Registration of CT, PET and additional MRI to the reference T1-weighted MRI | SimpleITK 2.5.6; six-degree-of-freedom rigid registration using Mattes mutual information, Powell optimization and geometry-based initialization with multiple candidate starting orientations |
| FreeSurfer version and imported outputs | FreeSurfer 8.2.0, used outside the application; orig.mgz, left and right pial surfaces, and aparc+aseg.mgz. Additional FreeSurfer surfaces and anatomical segmentations are loaded when available |
| Diffusion processing | dcm2niix 1.0.20260724 for DICOM conversion; SimpleITK 2.5.6 for optional rigid motion correction and image alignment; DIPY 1.12.1 for tensor and FA reconstruction and deterministic tractography |
| CT contact detection | Custom implementation using NumPy and SciPy for native-CT metal thresholding, connected-component analysis and lead grouping, followed by analyst review |
| Bipolar spatial assignment | Midpoint of the two assigned contacts in the reference MRI coordinate frame |
| Surface projection kernel and default distance | Normalized inverse-square-distance weighting within a user-defined neighborhood; default maximum projection distance, 15 mm |
| Analysis table import | openpyxl 3.1.5; Excel (.xlsx) tables |
| Color mapping | Matplotlib 3.11.2 |
| Video encoding | imageio-ffmpeg 0.6.0 with FFmpeg 7.1; H.264-encoded MP4 |
| Export formats and project storage | PNG images; MP4 videos; CSV contact tables containing anatomical labels and coordinates; JSON contact-coordinate records; project bundles containing JSON manifests, NIfTI images and masks, and NPZ numerical arrays |
| Operating systems tested | Windows 11 Enterprise, 64-bit; documented environment build 22631 |
| Version / commit | Prototype 1.0 RC20; software version 1.0.0rc20. Public release commit: https://github.com/yutanoue0712-bit/Omni_iEEG_Planner (doi:10.5281/zenodo.23028674) |
| License | MIT License |